\documentclass[
 reprint,
 nofootinbib,
 aps,
 prl,
]{revtex4-2}

\usepackage{amsmath,amssymb,mathrsfs,multirow}
\usepackage{ORCIDinREVTeX}
\usepackage{graphicx}
\graphicspath{{figures/}}
\usepackage{subfig}
\usepackage{dcolumn}
\usepackage{bm,slashed}
\usepackage[colorlinks=true,linkcolor=blue,urlcolor=blue,citecolor=blue]{hyperref}

\allowdisplaybreaks
\begin{document}

\title{The force of attraction between nucleons due to vacuum fluctuation}

\author{Anupam Ghosh}
\orcid{0000-0003-4163-4491}
\email{anupamg@rnd.iitg.ac.in}
\email{anupamghosh993@gmail.com}
\affiliation{Department of Physics, Indian Institute of Technology Guwahati, North Guwahati, 781039, Assam, India}

\begin{abstract}
We investigate quantum vacuum interactions arising from the zero-point fluctuations of a spatially confined massive scalar field. Deriving analytical expressions for the planar interaction energy and vacuum pressure, we identify a fundamental transition from the canonical power-law scaling of massless fields to a distinct quantum saturation regime. We prove that in the macroscopic limit, where the boundary separation far exceeds the field’s Compton wavelength ($a \gg \lambda_c$), the interaction energy does not vanish; instead, it asymptotes to a persistent constant, $-\hbar c / (24\pi \lambda_c^3)$. This reveals a cohesive zero-point energy reservoir inherent to massive vacua. Applying this formalism to the femtometer scale of the deuteron ($^2$H) nuclei, we demonstrate that confining massive pion fluctuations generates an attractive force between nucleons. 
\end{abstract}

\maketitle

\section{Introduction} 

The realization that the quantum vacuum is not an inert void but a highly dynamical medium governed by zero-point fluctuations is one of the cornerstone triumphs of quantum field theory \cite{Milonni:1994xx,Milton:2001yy}. The most direct macroscopic manifestation of this phenomenon is the Casimir effect \cite{Casimir:1948dh,Casimir:1947kzi}, where the spatial confinement of vacuum modes between boundaries alters the zero-point energy and induces an observable pressure. Over decades, high-precision laboratory measurements of the electromagnetic Casimir effect have established stringent constraints on fundamental physics, including limits on non-Newtonian gravity and hypothetical long-range interactions \cite{Lamoreaux:1999nw,Decca:2007yb,Bordag:2001qi,Decca:2003td,Decca:2005yk,Decca:2007jq}. Crucially, this boundary dependence of the quantum vacuum is not restricted to quantum electrodynamics; it extends universally across quantum field theory, playing an essential role in the structure of hadronic matter and the dynamics of massive fields.

The electromagnetic Casimir effect, mediated by massless $U(1)$ gauge bosons, exhibits pure power-law scaling dictated by boundary geometry~\cite{Asorey:2013wca,Ambjorn:1981xw}. The massive-field case is more intricate. Boundary-induced vacuum interactions are governed by the field's differential operator, with spatial boundary conditions acting as an effective confining potential~\cite{Bordag:2001qi}. The boundary-induced vacuum interaction, whether evaluated canonically via zero-point mode summation or extracted from the path-integral effective action, fundamentally differs from the massless case with the introduction of a mass gap $m$. This parameter explicitly breaks the conformal symmetry of the vacuum and imposes an intrinsic length scale: the Compton wavelength $\lambda_c = \hbar/mc$~\cite{Callan:1970ze}. Consequently, the interaction energy shifts from simple inverse-power laws to complex, scale-dependent formalisms. Extracting finite physical observables, such as vacuum pressure or energy density, requires regularization and handling ultraviolet divergences. The general structure of these infinities is managed by combining the zeta-functional regularization with heat kernel expansions~\cite{Dowker:1975tf,Hawking:1976ja,DeWitt:1975ys,Blau:1988kv,Elizalde:1989dd,Elizalde:1994gf,Elizalde:1995hck,Bordag:1996ma}. Understanding this massive boundary-induced vacuum pressure remains of profound interdisciplinary interest, since massive scalar fields govern dynamics across all physical energy scales. Within the Standard Model, they appear as the Higgs boson~\cite{Higgs:1964pj, Englert:1964et} and as pseudo-Goldstone bosons, such as the pion. Furthermore, many beyond-the-Standard-Model theories rigorously predict a vast spectrum of novel scalar degrees of freedom. These include axion-like particles introduced to resolve the strong CP problem~\cite{Peccei:1977hh,Peccei:1977ur}, as well as other hypothetical dark-sector scalars~\cite{Boehm:2003hm,Burgess:2000yq,Barbieri:2006dq}.

In this work, we provide a rigorous analytical derivation of the interaction energy and force density between two parallel plates resulting from massive scalar field fluctuations, assuming Dirichlet boundary conditions. Evaluated in the zero-temperature limit, our derivation uncovers a distinct quantum saturation regime. We demonstrate that as the boundary separation $a$ significantly exceeds $\lambda_c$, conformal symmetry breaking prevents the interaction energy density from asymptotically vanishing. Instead, it asymptotes to a persistent constant, acting as an intrinsic energy reservoir.

To demonstrate the phenomenological impact of this result on the nuclear femtometer scale, we utilize the massive pion field as an exemplary scalar. Recognizing that constituent nucleons undergo continuous quantum fluctuations~\cite{Perkins:1982xb} (e.g., $p \longleftrightarrow n + \pi^+$), we treat nucleons as spatial boundaries that restrict the surrounding pion vacuum. Considering the deuteron ($^{2}$H) as a baseline test, we approximate the energetic transition from isolated free nucleons (an unrestricted vacuum) to a bound composite state (a restricted vacuum). Nuclear stability requires a persistent attractive force to balance the strongly repulsive short-range core ($a \lesssim 0.5$ fm) that arises from Pauli exclusion among overlapping constituent quarks. A review on nuclear forces can be found in reference~\cite{Epelbaum:2008ga} and references therein. Our results reveal that the required attractive reservoir emerges naturally from the spatial confinement of the pion field. This suggests that the attractive strong nuclear interaction can be understood as a geometric manifestation of massive zero-point energy. In the following section, we derive exact analytical expressions for the interaction energy density and vacuum pressure of a massive scalar field subject to spatial confinement.

\section{Force between parallel plates due to vacuum fluctuations}
Consider two parallel square plates, each of area $L^2$, positioned parallel to the $xy$-plane with an inter-plate spacing $a$ along the $z$-axis. The physical interaction energy is defined as the difference in zero-point energy between two distinct configurations,
\begin{equation}
    \delta E = \left( \sum_n \frac{1}{2} \hbar \omega_n \right)_{I} - \left( \sum_n \frac{1}{2} \hbar \omega_n \right)_{II}\,.
    \label{eq.delE-1}
\end{equation}
Configuration $I$ describes the restricted vacuum at a finite plate separation $a$, which quantizes the normal wave number $k_z$ into discrete modes. Configuration $II$ defines the asymptotic reference state ($a \gg \lambda_c$), where $\lambda_c$ is the reduced Compton wavelength of the scalar field. In Configuration $II$, the boundary restrictions vanish entirely, and $k_z$ spans an unrestricted continuum. Although the individual zero-point energy sums in equation\,\ref{eq.delE-1} are ultraviolet divergent, the difference between these two configurations identically cancels the divergences, yielding a finite physical observable.

We consider a massive scalar field $\Phi$ of mass $m$, subject to Dirichlet boundary conditions at the plates ($\Phi|_{z=0,a}=0$). In $(3+1)$-dimensional Minkowski spacetime, the canonical mode expansion of the quantized field takes the form
\begin{align}
    \hat{\Phi}(t, x, y, z) &= \sum_{n=1}^{\infty} \int \frac{dk_x dk_y}{(2\pi)^2} \frac{1}{\sqrt{2E_n}} \left[ \hat{a}_{n, k_x, k_y} u_n(t, x, y, z) \right. \nonumber \\
    &\quad + \left. \hat{a}^\dagger_{n, k_x, k_y} u^*_n(t, x, y, z) \right] ,
\end{align}
where $\hat{a}$ and $\hat{a}^\dagger$ denote the bosonic annihilation and creation operators. The corresponding mode functions satisfying the Dirichlet boundary conditions are given by
\begin{equation}
    u_n(t, x, y, z) \propto \sin\left( \frac{n \pi z}{a} \right) \exp\left[ i(k_x x + k_y y - \omega_n t) \right].
\end{equation}
This geometric confinement quantizes the normal wave number as $k_z = n \pi / a$. The mode index takes positive integer values ($n \in \{1, 2, 3, \dots\}$), and the $n=0$ state yields a trivial, unphysical solution where the field vanishes entirely.
The energy dispersion relation for the massive field becomes
\begin{equation}
    \omega_n = \frac{E_n}{\hbar} = c \left[  k_x^2 + k_y^2 + \frac{n^2\pi^2}{a^2}  + \frac{m^2 c^2}{\hbar^2} \right]^{1/2}.
\end{equation}
The momenta $k_x$ and $k_y$ can be considered continuous variables for large $L$. Integrating over the continuous transverse modes using polar coordinates in the $(k_x, k_y)$ plane, the unregularized zero-point energy for configuration $I$ becomes,
\begin{align}
    \left( \sum_n \frac{1}{2} \hbar \omega_n \right)_{I} &= \frac{\hbar c L^2}{4\pi} \sum_{n=1}^{\infty} \int_{0}^{\infty} \sqrt{ \frac{n^{2}\pi^{2}}{a^{2}} + x^{2} + \frac{m^{2}c^{2}}{\hbar^{2}}}\; x\, dx\,.
\end{align}

In configuration II, the discrete summation over $n$ can be replaced with the integral over the continuum variable $k_z$. The interaction energy becomes,
\begin{align}
    \delta E &= \frac{\hbar c L^{2}}{4\pi} \bigg[ \sum_{n=1}^{\infty} \int_{0}^{\infty} \sqrt{ \frac{n^{2}\pi^{2}}{a^{2}} + x^{2} + \frac{m^{2}c^{2}}{\hbar^{2}} }\; x\,dx \nonumber \\
    &\qquad\quad - \int \int  \sqrt{ k_{z}^{2}+x^{2} + \frac{m^{2}c^{2}}{\hbar^{2}} }\; x\,dx \left(\frac{a}{\pi}\, dk_{z}\right) \bigg].
\end{align}
By introducing a dimensionless variable $u = \frac{a^2}{\pi^2} \left( x^2 + \frac{m^2c^2}{\hbar^2} \right)$, the interaction energy can be rewritten as
\begin{align}
 \delta E
 &=
 \frac{\hbar c L^{2}}{4\pi}
 \frac{\pi^{3}}{2a^{3}}
 \bigg[
 \sum_{n=1}^{\infty}
 \int_{u_{\min}}^{\infty}
 \sqrt{n^{2}+u}\,\, du
 \nonumber\\[4pt]
 &\qquad\quad
 -
 \int_{n=1}^{\infty}  \int_{u_{\min}}^{\infty}\sqrt{n^{2}+u}\,\, du \, dn \bigg],
 \label{eq.delE-2}
\end{align}
where $u_{\min} = \frac{a^{2} m^{2} c^{2}}{\pi^{2}\hbar^{2}}.$ 
We can express equation\,\ref{eq.delE-2} as,
\begin{equation}
    \delta E = \frac{\hbar c L^{2}}{4\pi} \frac{\pi^{3}}{2a^{3}} I(u_{\min})\,.
    \label{Eq:dE}
\end{equation}
Here, the functional $I(u_{\min})$ encodes the difference between discrete boundary modes and the unrestricted continuous vacuum limit,
\begin{equation}
    I(u_{\min}) = \sum_{n=1}^{\infty} \int_{u_{\min}}^{\infty} \sqrt{n^{2}+u}\; du \,- \int_{n=1}^{\infty}  \int_{u_{\min}}^{\infty}\sqrt{n^{2}+u}\; du \, dn\,.
\end{equation}
\subsection{Extraction of the Finite Physical Contribution.}
Because the constituent integrals of $I(u_{\min})$ diverge in the ultraviolet limit, we introduce an exponential regulator to the core integrand, $f(n) = \int_{u_{\min}}^{\infty} \sqrt{n^{2}+u}\,\, du$, yielding
\begin{equation}
    f_{\epsilon}(n) = \int_{u_{\min}}^{\infty} \sqrt{n^{2}+u}\,\, e^{-\epsilon (n^{2}+u)}\, du\,, \qquad \epsilon>0\,.
    \label{eq-regu}
\end{equation}
The regularized energy functional then becomes
\begin{equation}
    I_\epsilon(u_{\min}) = \sum_{n=1}^{\infty} f_\epsilon(n) - \int_{n=1}^{\infty} f_\epsilon(n) \, dn\,,
    \label{eq-Ieps}
\end{equation}
which exactly recovers $I(u_{\min})$ in the $\epsilon \to 0$ limit. Substituting $t = n^2 + u$ maps the integral in equation\,\ref{eq-regu} directly onto the upper incomplete gamma function \cite{Olver:2010ouy}, we obtain, 
\begin{equation}
    f_\epsilon(n) = \epsilon^{-3/2} \Gamma\left(\frac{3}{2}, \epsilon(n^2+u_{\min})\right).
\end{equation}
Using the standard series expansion $\Gamma(s, x) = \Gamma(s) - \frac{x^s}{s} + \mathcal{O}(x^{s+1})$, we find that the $\epsilon \to 0$ limit yields a universal divergent term, $\Gamma(3/2)\epsilon^{-3/2}$. Because this divergence is independent of $n$, it identically cancels when taking the difference in equation\,\ref{eq-Ieps}. The remaining finite part of $f_\epsilon(n)$ as $\epsilon \to 0$ is simply
\begin{equation}
    F(n) = -\frac{2}{3}(n^2 + u_{\min})^{3/2}\,.
\end{equation}
Applying the Euler-Maclaurin summation formula \cite{Olver:2010ouy} to equation\,\ref{eq-Ieps} in this limit yields
\begin{equation}
I(u_{\min}) = \frac{1}{2}F(1) - \frac{1}{12}F'(1) + \frac{1}{720}F'''(1) + \cdots
\label{eq-Iu}
\end{equation}
where the prime denotes the derivative with respect to $n$. Evaluating these derivatives at $n=1$, we obtain
\begin{align}
    F(1) &= -\frac{2}{3}(1+u_{\min})^{3/2}\,, \\
    F'(1) &= -2\sqrt{1+u_{\min}}\,, \\
    F'''(1) &= -6(1+u_{\min})^{-1/2} + 2(1+u_{\min})^{-3/2}\,.
\end{align}
Substituting these into equation\,\ref{eq-Iu}, the functional evaluates to
\begin{equation}
    I(u_{\min}) = -\frac{(1+u_{\min})^{3/2}}{3} + \frac{\sqrt{1+u_{\min}}}{6} - \frac{1}{120\sqrt{1+u_{\min}}} + \dots
    \label{Eq:Int-final}
\end{equation}
\section{Final Result and Discussion}
Rewriting $u_{\min}$ explicitly in terms of the field's reduced Compton wavelength $\lambda_c = \hbar/mc$ gives $u_{\min} = \frac{1}{\pi^2} (a/\lambda_c)^2$. Combining equations~\ref{Eq:dE} and \ref{Eq:Int-final}, we obtain the interaction energy,
\begin{widetext}
\begin{align}
\frac{\delta E}{L^{2}}
& =-\frac{\hbar c\, \pi^{2}}{8a^{3}}
\bigg[ \frac{(1+u_{\min})^{3/2}}{3} - \frac{\sqrt{1+u_{\min}}}{6} + \frac{1}{120\sqrt{1+u_{\min}}} + \dots \bigg]
\nonumber\\[4pt]
\qquad\quad
&\approx- \frac{\hbar c\, \pi^{2}}{960\,a^{3}} \frac{21 + 20\, u_{\min} (3+2\,u_{\min})}{\sqrt{1+u_{\min}}}
 \, .
\label{Eq-E-final}
\end{align}
The negative interaction energy signifies an attractive, cohesive vacuum force between the boundaries. The corresponding vacuum pressure (force per unit area) is derived from the negative spatial gradient of the planar energy density, $F/L^{2} = -\frac{d}{da} (\delta E/L^{2})$, which yields
\begin{align}
    \frac{F}{L^{2}}
    & =  - \frac{\hbar c \pi^{2}}{8a^{4}}
    \bigg[ \frac{\sqrt{1+u_{\min}}}{2} + \frac{u_{\min}}{6\sqrt{1+u_{\min}}} + \frac{1}{40\sqrt{1+u_{\min}}}   + \mathcal{O}\bigg( \frac{u_{\min}}{(1+u_{\min})^{3/2}} \bigg)  \bigg]
  \nonumber\\[4pt]
  \qquad\quad
   &=    - \frac{\hbar c \pi^{2}}{960\,a^{4}}
   \bigg[ \frac{63+80\,u_{\min}}{\sqrt{1+u_{\min}}} + \mathcal{O}\bigg( \frac{u_{\min}}{(1+u_{\min})^{3/2}} \bigg)  \bigg]\, .
    \label{Eq-F-final}
\end{align}
\end{widetext}

\begin{figure*}[!htb]
    \centering
        \includegraphics[width=0.48\textwidth]{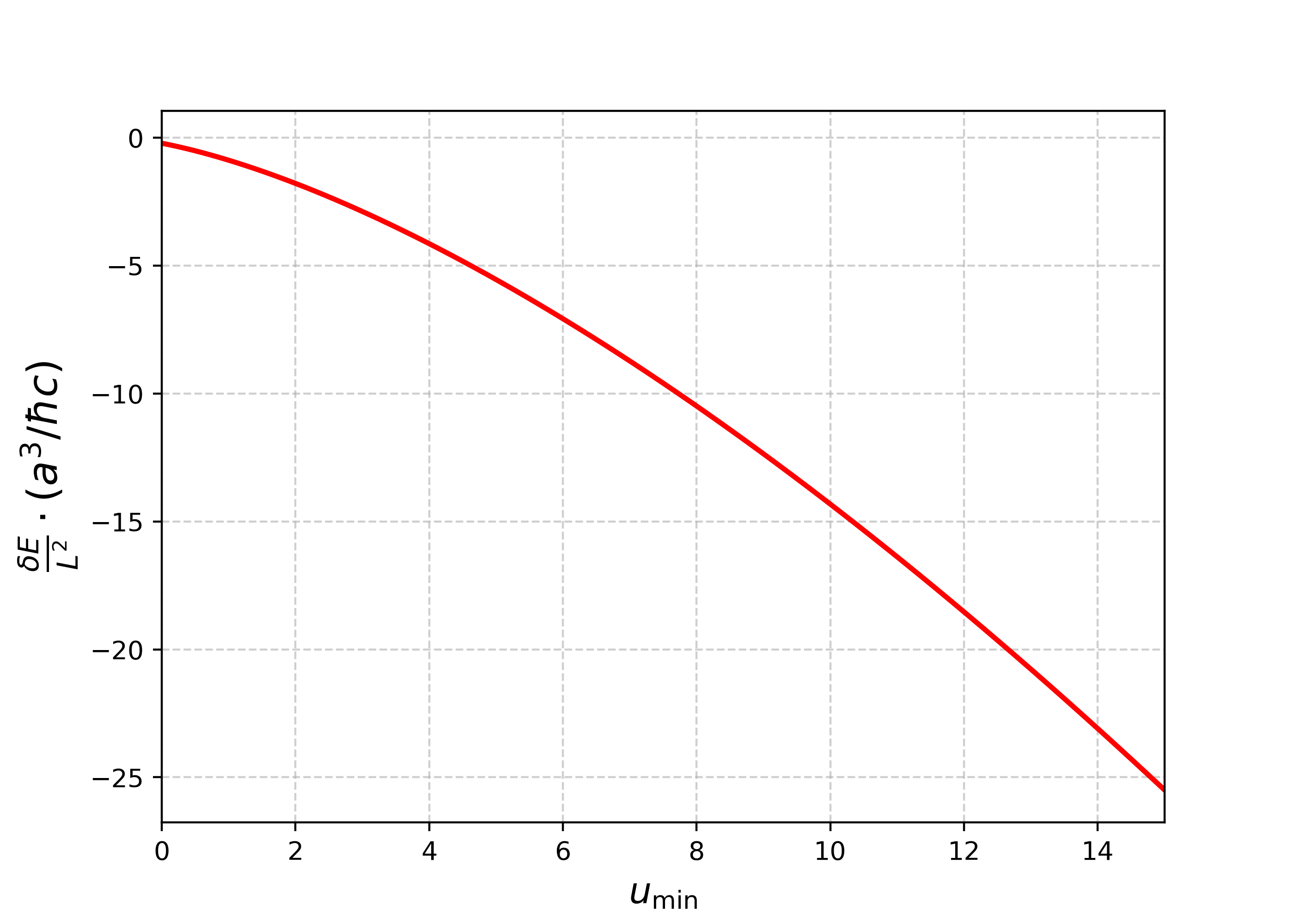}
            \quad
    \includegraphics[width=0.48\textwidth]{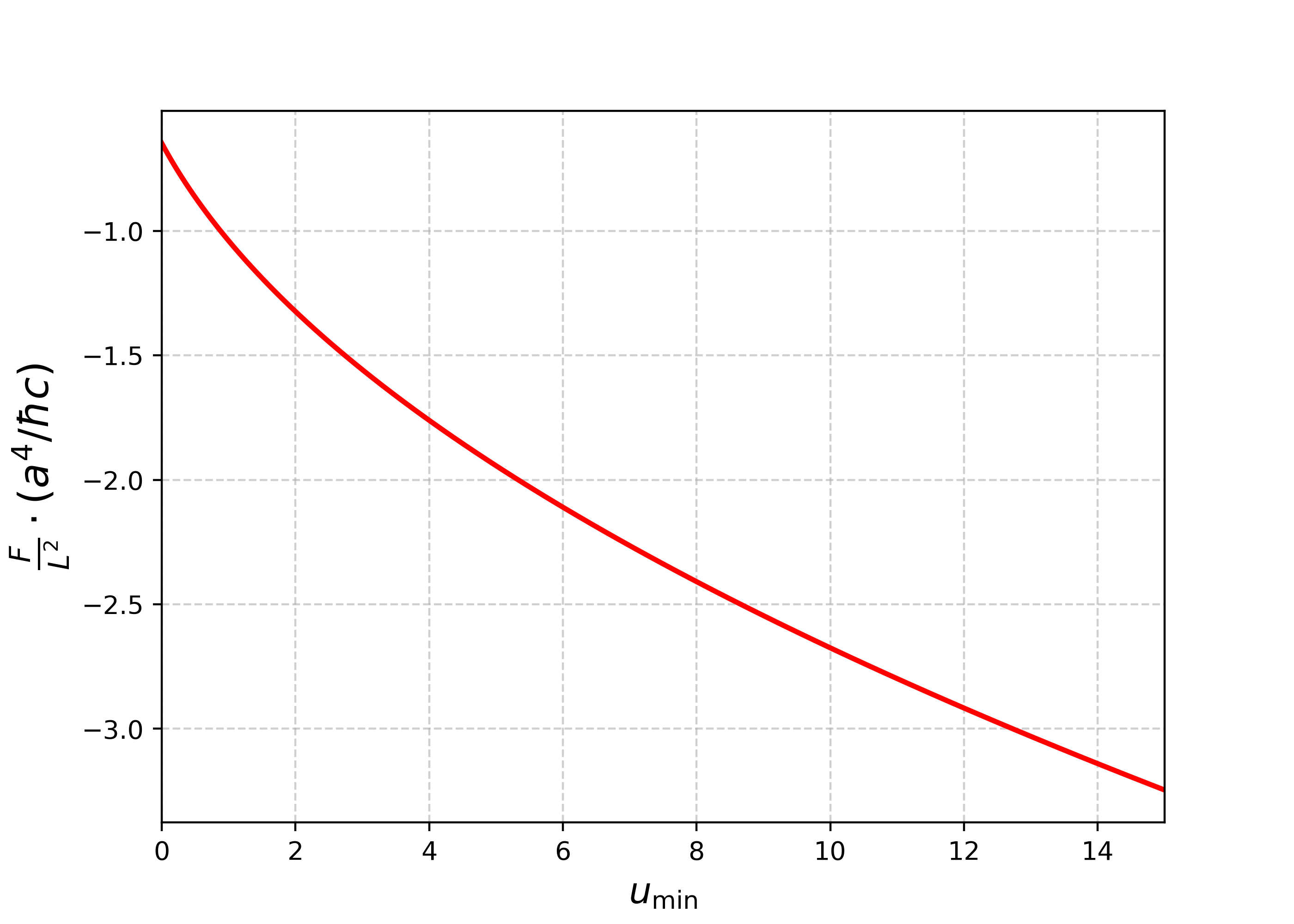}
    \caption{Variation of the interaction energy and force per unit area as functions of the dimensionless variable $u_{\min} =\frac{1}{\pi^2} (a/\lambda_c)^2$. }
    \label{fig:dimensionless}
\end{figure*}

\begin{figure*}[!htb]
    \centering
        \includegraphics[width=0.48\textwidth]{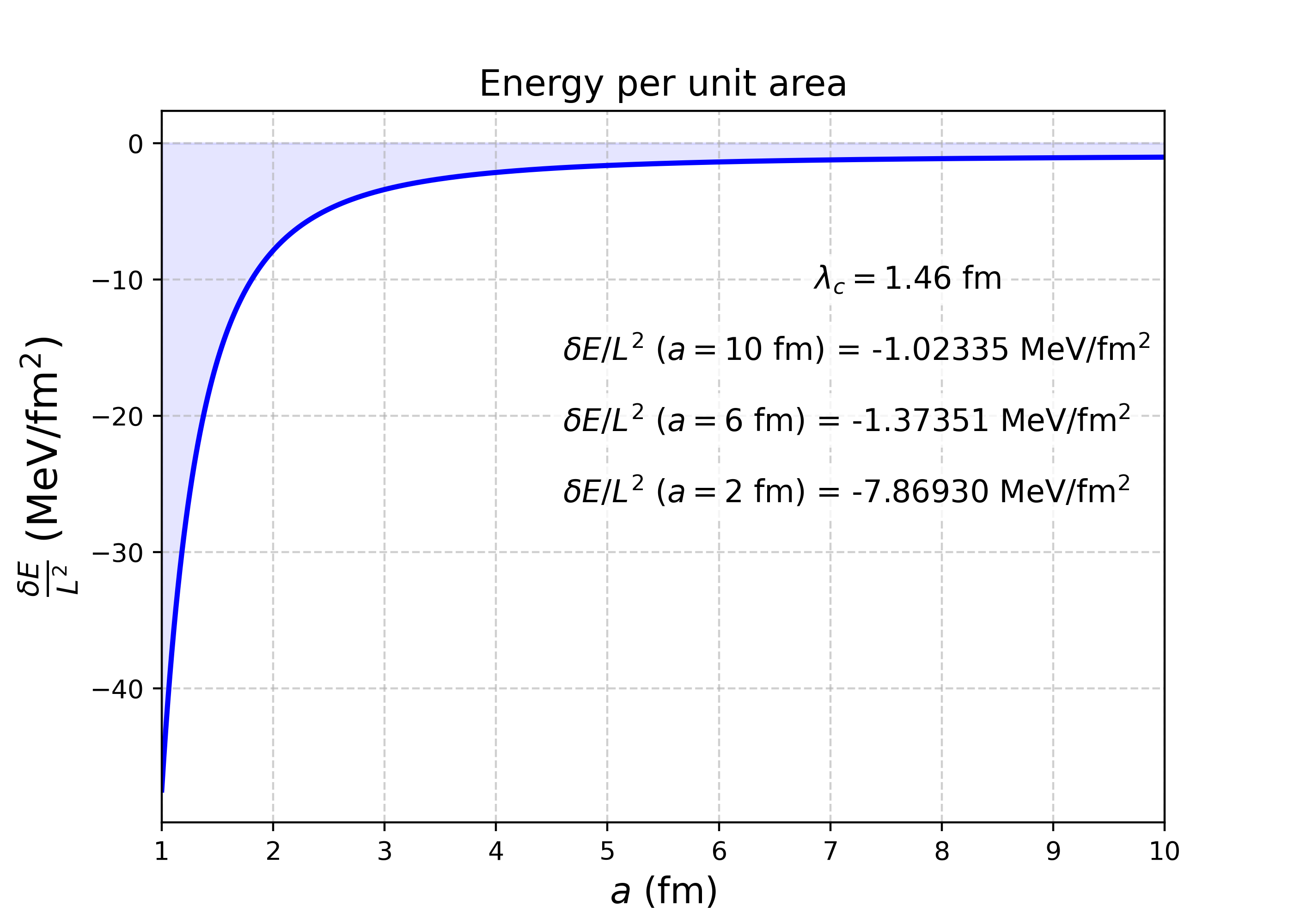}
    \quad
    \includegraphics[width=0.48\textwidth]{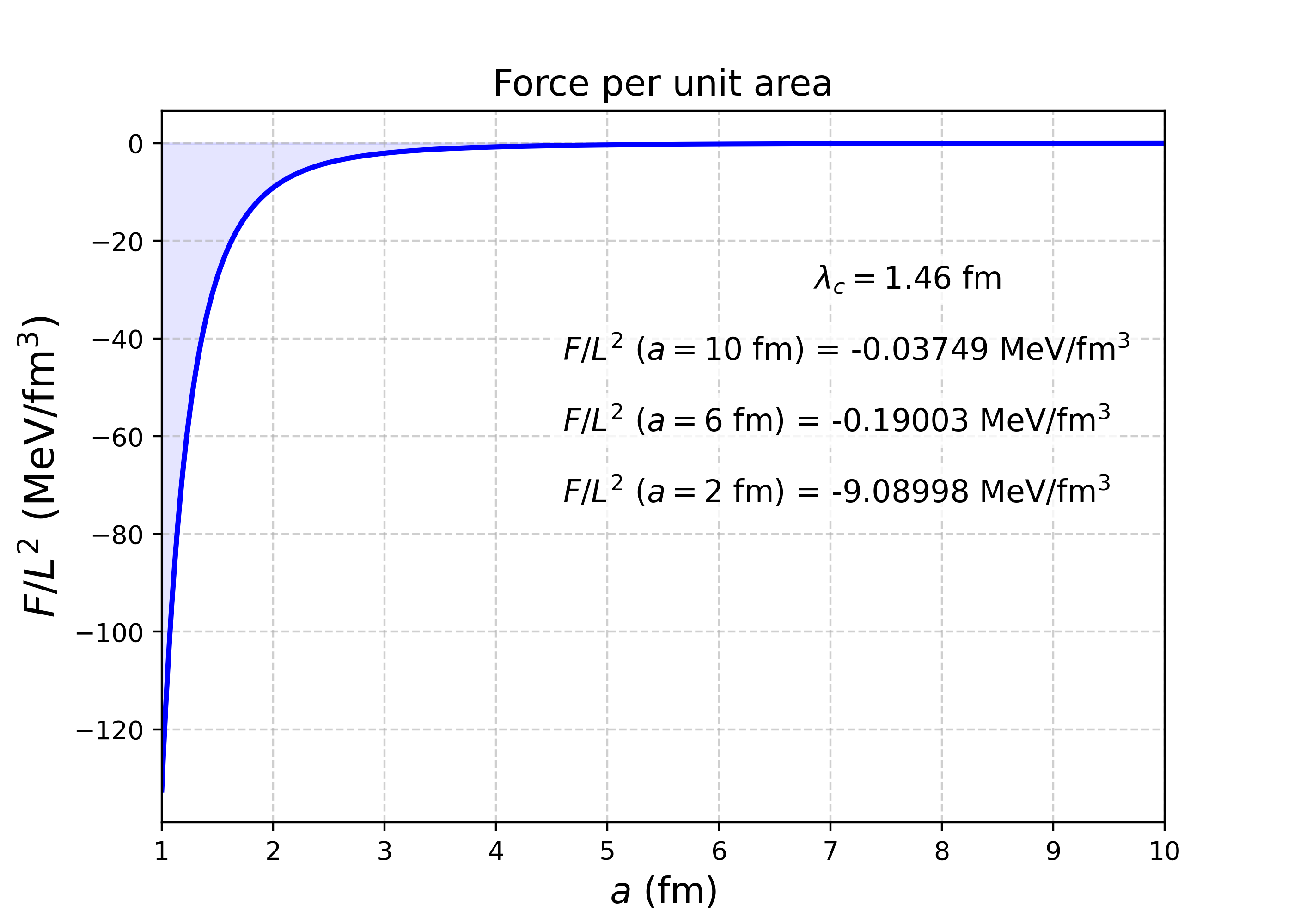}
    \caption{Variation of the interaction energy per unit area (left, in MeV/fm$^2$) and force per unit area (right, in MeV/fm$^3$) as a function of separation $a$ (in fm), evaluated for $\lambda_c = 1.46$~fm. The values of these quantities corresponding to $a=2,\,6,\, 10$ fm are also given.}
    \label{fig:plots}
\end{figure*}

\begin{figure*}[!htb]
    \centering
        \includegraphics[width=0.48\textwidth]{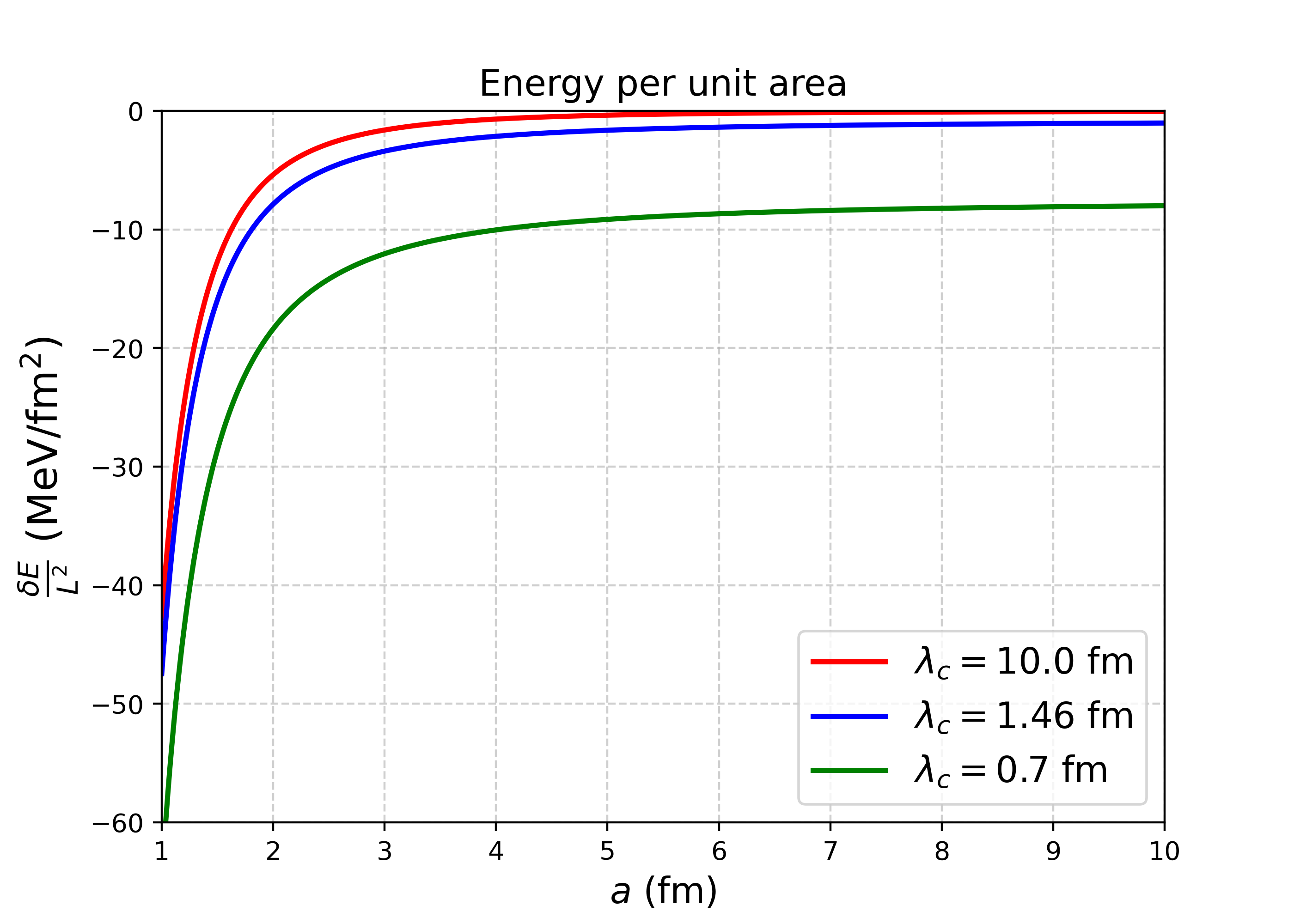}
            \quad
    \includegraphics[width=0.48\textwidth]{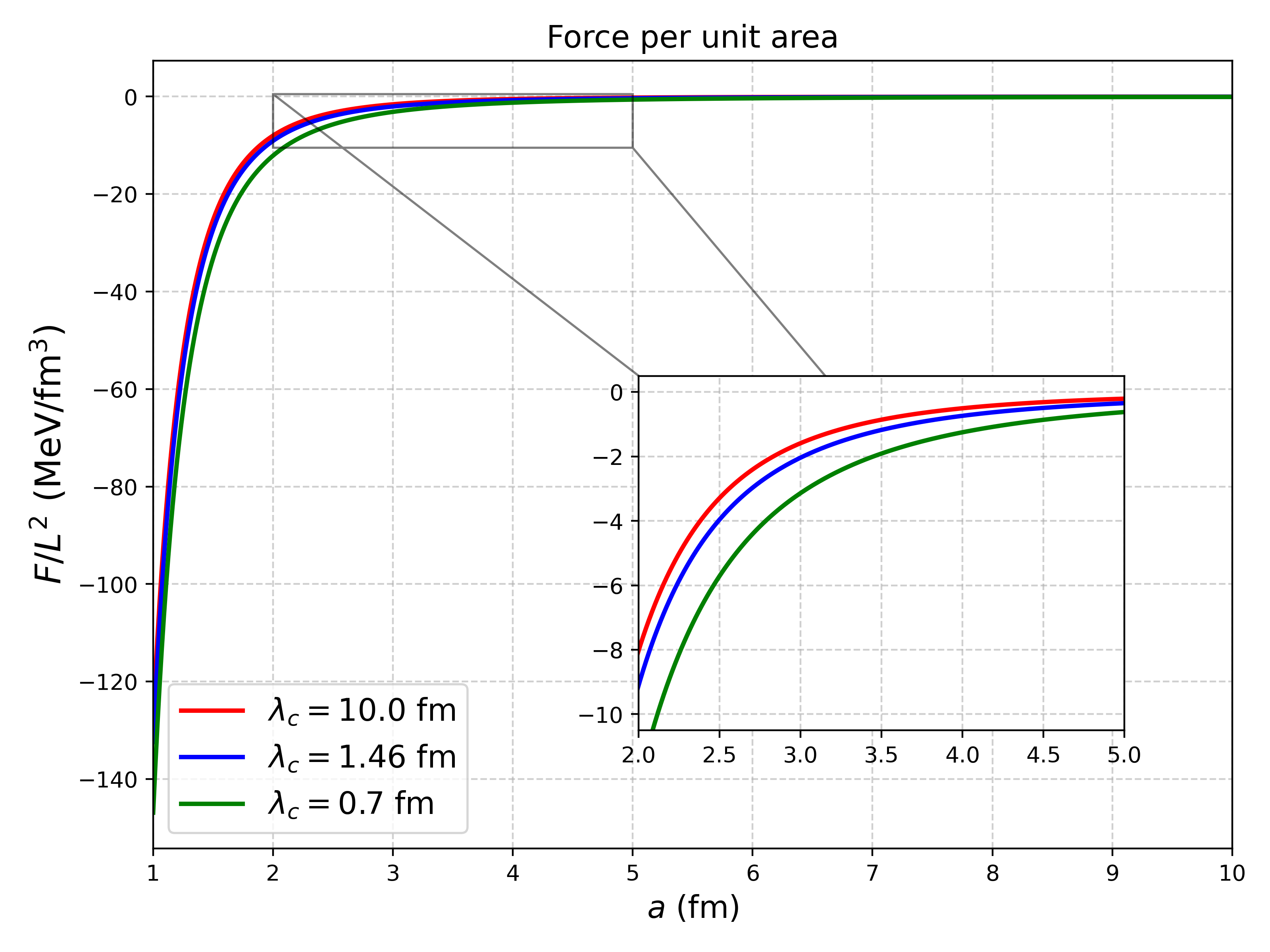}
    \caption{Same as figure\,\ref{fig:plots} but for different $\lambda_c$ values.}
    \label{fig:plots2}
\end{figure*}
Figure~\ref{fig:dimensionless} presents the behavior of the interaction energy and force per unit area as functions of the dimensionless scale $u_{\min}$. Setting the Compton wavelength to that of the pion ($\lambda_c = 1.46$ fm), figure~\ref{fig:plots} illustrates the corresponding interaction energy per unit area and vacuum pressure as functions of the separation distance $a$. At short distances, both quantities grow rapidly, driven by their dominant $1/a^3$ and $1/a^4$ scaling behaviors, respectively. Figure~\ref {fig:plots2} then investigates the variation of the energy density and pressure with respect to $a$ across different values of $\lambda_c$. Decreasing $\lambda_c$, which corresponds to a larger scalar field mass, shifts both the vacuum energy and pressure downward, enhancing the magnitude of the interaction at any fixed separation.
\textbf{Conformal limit.} In the massless limit ($m \to 0$, $u_{\min} \to 0$), conformal symmetry is restored, and the interaction energy simplifies to
\begin{equation}
 \frac{\delta E}{L^{2}} \approx -\frac{21\,\hbar c \,\pi^{2} }{960\, a^{3}} \, , \qquad\quad \frac{F}{L^{2}} \approx -\frac{21\,\hbar c\, \pi^{2} }{320\, a^{4}}\, .
\end{equation}
Therefore, in this limit, our formalism explicitly recovers the standard massless Casimir effect, restoring the canonical $1/a^3$ energy scaling alongside the associated $1/a^4$ vacuum pressure.
\textbf{Massive saturation limit.} The central theoretical result of this work emerges in the long-range limit. When the boundary separation significantly exceeds the intrinsic Compton scale ($a \gg \lambda_c$), equations~\ref{Eq-E-final} and \ref{Eq-F-final} asymptotically reduce to 
\begin{align}
    &\frac{\delta E}{L^{2}} \approx - \frac{\hbar c }{24\pi \lambda_c^{3}} \big[ 1 + \pi^2 \frac{\lambda_c^2}{a^2} \big]   \label{eq:asym}  \, ,
    \\[4pt]
    \qquad\quad
    & \frac{F}{L^{2}} \approx - \frac{\hbar c \,\pi }{12\,a^3 \lambda_c}\, .
\end{align}
This constitutes a \textit{saturation phenomenon}. Unlike the electromagnetic Casimir effect, which vanishes at infinite separation, the bounded massive field retains a persistent energetic signature. The leading term of the interaction energy density, $-\frac{\hbar c}{24\pi \lambda_c^{3}}$, acts as a constant offset driven by spatial confinement. Restricting the massive vacuum enforces a zero-point energy deficit that serves as a cohesive reservoir. Concurrently, the vacuum pressure transitions to an inverse-cubic scaling, $\propto 1/(a^3 \lambda_c)$. This algebraic tail fundamentally alters long-range boundary dynamics, departing from the standard $1/a^4$ characteristic of massless fields.

\vspace{1em}
\noindent \textbf{Implications for the Deuteron Nucleus.}
Modern nuclear physics predominantly describes the nucleon-nucleon interaction through chiral effective field theories, wherein long-range forces are mediated by virtual pion exchanges and unresolved short-range dynamics are parameterized by localized contact interactions \cite{Weinberg:1990rz,Epelbaum:2008ga,Machleidt:2011zz}. However, evaluating the interaction energy arising from vacuum fluctuations offers a useful complementary perspective. Nevertheless, an attractive force between nucleons is needed for nuclear stability.

To establish this correspondence, we interpret the nucleons as physical boundaries that spatially restrict the massive pion vacuum ($m_{\pi} \approx 135$ MeV/$c^2$, $\lambda_c \approx 1.46$ fm). The asymptotic reference state represents non-interacting, isolated nucleons enveloped by an unrestricted vacuum. When brought into close proximity (finite $a$), the nucleons geometrically confine the pion field, thereby quantizing the allowed momentum modes. The resulting negative interaction energy, $\delta E$, effectively acts as a cohesive force. To separate the nucleons back to their free states, external energy must be supplied to restore the excluded vacuum modes. 

For a quantitative estimate, we apply our formalism to the deuteron ($^{2}$H) nucleus. We estimate the total induced vacuum energy by scaling the planar energy density, $\delta E/L^2$, by the effective interaction cross-section of the nucleons. Assuming a standard nucleon charge radius of $R_n \approx 0.85\,\text{fm}$ \cite{Mohr:2024kco}, the effective geometrical area is $A_{\text{eff}} = \pi R_{n}^{2}$. The total theoretical interaction energy is then estimated as $E_{\text{theory}}(a) = (\delta E/L^2) \cdot A_{\text{eff}}$. Using the energy density values obtained from equation~\ref{Eq-E-final} (also quoted in figure\,\ref{fig:plots}) for $\lambda_c = 1.46$ fm, we find that at a separation of $a = 2.0$ fm, the total bound state energy evaluates to $E_{\text{theory}} \approx -17.86$ MeV. At an extended separation of $a = 6.0$ fm, this energy reduces to $E_{\text{theory}} \approx -3.12$ MeV.

It is important to note a geometric caveat: modeling spherical nucleons as infinite planar boundaries inherently overestimates the interaction, as highly curved surfaces allow vacuum modes to diffract into the exterior unconfined region. Consequently, the derived $-17.86$ MeV serves as a theoretical upper bound.  In nuclei, this attractive vacuum force could lead to structural collapse, which is prevented by a strongly repulsive short-range force ($a \lesssim 0.5$ fm), mediated by the Pauli exclusion principle. The vacuum energy reservoir identified here can provide a sufficient cohesive background to balance this short-range repulsion, contributing to the stabilization of the nucleus.
\section{Conclusion}
We have derived analytical expressions for the interaction energy and vacuum pressure of a confined massive scalar field. Our results establish that field mass fundamentally alters long-range quantum vacuum dynamics. In the macroscopic regime ($a \gg \lambda_c$), the planar interaction energy asymptotes to a constant proportional to $\lambda_c^{-3}$, while the vacuum pressure exhibits an inverse-cubic tail, $\propto 1/(a^3\lambda_c)$. This saturation regime contrasts sharply with the Casimir effect of massless fields.

Phenomenologically, this saturation mechanism offers a direct geometric perspective on nuclear cohesion. Modeling nucleons as boundaries for the pion field demonstrates that confining zero-point modes generates a substantial inward vacuum pressure. While diffractive leakage around spherical boundaries renders the planar calculation a strict upper bound, the resulting vacuum reservoir remains substantial. This attractive background helps balance short-range Pauli repulsion, complementing standard chiral effective field theory models of nuclear equilibrium.

Furthermore, the analytical scaling behavior established in this work can have direct implications for macroscopic metrology. Precision torsion-balance and Casimir force experiments routinely probe the sub-millimeter regime to search for hypothetical massive scalars and axion-like particles \cite{Adelberger:2003zx, Decca:2007yb}. In these high-sensitivity setups, rigorously accounting for all boundary-induced vacuum backgrounds is crucial. The inverse-cubic pressure scaling and massive saturation limits derived in our formalism can provide some useful insights for modeling the Casimir-like signatures of these beyond-the-Standard-Model candidates. Therefore, investigating the properties of restricted massive vacua can reveal deep physical connections across vastly different energy scales. 
\section{ACKNOWLEDGMENTS}
The author gratefully acknowledges Prof. Meduri C. Kumar for reading the manuscript and providing helpful comments and corrections. 
\bibliographystyle{apsrev4-2}
\bibliography{ref}

@book{Milonni:1994xx,
  title={The Quantum Vacuum: An Introduction to Quantum Electrodynamics},
  author={Milonni, P.W.},
  isbn={9780124980808},
  lccn={93029780},
  url={https://books.google.co.in/books?id=P83vAAAAMAAJ},
  year={1994},
  publisher={Elsevier Science}
}

@book{Milton:2001yy,
    author = "Milton, K. A.",
    title = "{The Casimir effect: Physical manifestations of zero-point energy}",
    doi = "10.1142/4505",
    isbn = "978-981-02-4397-5, 978-1-281-95620-0, 978-981-4492-50-8, 978-981-281-052-6",
    year = "2001"
}

@article{Bordag:2001qi,
    author = "Bordag, Michael and Mohideen, U. and Mostepanenko, V. M.",
    title = "{New developments in the Casimir effect}",
    eprint = "quant-ph/0106045",
    archivePrefix = "arXiv",
    doi = "10.1016/S0370-1573(01)00015-1",
    journal = "Phys. Rept.",
    volume = "353",
    pages = "1--205",
    year = "2001"
}

@book{Perkins:1982xb,
    author = "Perkins, D. H.",
    title = "{Introduction to high energy physics}",
    isbn = "978-0-521-62196-0",
    year = "1982"
}

@article{Casimir:1948dh,
    author = "Casimir, H. B. G.",
    title = "{On the attraction between two perfectly conducting plates}",
    journal = "Indag. Math.",
    volume = "10",
    number = "4",
    pages = "261--263",
    year = "1948"
}

@article{Casimir:1947kzi,
    author = "Casimir, H. B. G. and Polder, D.",
    title = "{The Influence of retardation on the London-van der Waals forces}",
    doi = "10.1103/PhysRev.73.360",
    journal = "Phys. Rev.",
    volume = "73",
    pages = "360--372",
    year = "1948"
}

@article{Lamoreaux:1999nw,
    author = "Lamoreaux, S. K.",
    title = "{Resource letter CF-1: Casimir force}",
    doi = "10.1119/1.19138",
    journal = "Am. J. Phys.",
    volume = "67",
    pages = "850--861",
    year = "1999"
}

@article{Decca:2003td,
    author = "Decca, R. S. and Fischbach, E. and Klimchitskaya, G. L. and Krause, D. E. and Lopez, D. L. and Mostepanenko, V. M.",
    title = "{Improved tests of extra dimensional physics and thermal quantum field theory from new Casimir force measurements}",
    eprint = "hep-ph/0310157",
    archivePrefix = "arXiv",
    doi = "10.1103/PhysRevD.68.116003",
    journal = "Phys. Rev. D",
    volume = "68",
    pages = "116003",
    year = "2003"
}

@article{Decca:2005yk,
    author = "Decca, R. S. and Lopez, D. and Fischbach, E. and Klimchitskaya, G. L. and Krause, D. E. and Mostepanenko, V. M.",
    title = "{Precise comparison of theory and new experiment for the Casimir force leads to stronger constraints on thermal quantum effects and long-range interactions}",
    eprint = "quant-ph/0503105",
    archivePrefix = "arXiv",
    doi = "10.1016/j.aop.2005.03.007",
    journal = "Annals Phys.",
    volume = "318",
    pages = "37--80",
    year = "2005"
}

@article{Decca:2007yb,
    author = "Decca, R. S. and Lopez, D. and Fischbach, E. and Klimchitskaya, G. L. and Krause, D. E. and Mostepanenko, V. M.",
    title = "{Tests of new physics from precise measurements of the Casimir pressure between two gold-coated plates}",
    eprint = "hep-ph/0703290",
    archivePrefix = "arXiv",
    doi = "10.1103/PhysRevD.75.077101",
    journal = "Phys. Rev. D",
    volume = "75",
    pages = "077101",
    year = "2007"
}

@article{Decca:2007jq,
    author = "Decca, R. S. and Lopez, D. and Fischbach, E. and Klimchitskaya, G. L. and Krause, D. E. and Mostepanenko, V. M.",
    title = "{Novel constraints on light elementary particles and extra-dimensional physics from the Casimir effect}",
    eprint = "0706.3283",
    archivePrefix = "arXiv",
    primaryClass = "hep-ph",
    doi = "10.1140/epjc/s10052-007-0346-z",
    journal = "Eur. Phys. J. C",
    volume = "51",
    pages = "963--975",
    year = "2007"
}

@article{Epelbaum:2008ga,
    author = "Epelbaum, Evgeny and Hammer, Hans-Werner and Meissner, Ulf-G.",
    title = "{Modern Theory of Nuclear Forces}",
    eprint = "0811.1338",
    archivePrefix = "arXiv",
    primaryClass = "nucl-th",
    reportNumber = "HISKP-TH-08-18, FZJ-IKP-TH-2008-20",
    doi = "10.1103/RevModPhys.81.1773",
    journal = "Rev. Mod. Phys.",
    volume = "81",
    pages = "1773--1825",
    year = "2009"
}

@article{Asorey:2013wca,
    author = "Asorey, M. and Munoz-Castaneda, J. M.",
    title = "{Attractive and Repulsive Casimir Vacuum Energy with General Boundary Conditions}",
    eprint = "1306.4370",
    archivePrefix = "arXiv",
    primaryClass = "hep-th",
    doi = "10.1016/j.nuclphysb.2013.06.014",
    journal = "Nucl. Phys. B",
    volume = "874",
    pages = "852--876",
    year = "2013"
}

@article{Higgs:1964pj,
    author = "Higgs, Peter W.",
    editor = "Taylor, J. C.",
    title = "{Broken Symmetries and the Masses of Gauge Bosons}",
    doi = "10.1103/PhysRevLett.13.508",
    journal = "Phys. Rev. Lett.",
    volume = "13",
    pages = "508--509",
    year = "1964"
}

@article{Englert:1964et,
    author = "Englert, F. and Brout, R.",
    editor = "Taylor, J. C.",
    title = "{Broken Symmetry and the Mass of Gauge Vector Mesons}",
    doi = "10.1103/PhysRevLett.13.321",
    journal = "Phys. Rev. Lett.",
    volume = "13",
    pages = "321--323",
    year = "1964"
}

@article{Peccei:1977hh,
    author = "Peccei, R. D. and Quinn, Helen R.",
    title = "{CP Conservation in the Presence of Instantons}",
    reportNumber = "ITP-568-STANFORD",
    doi = "10.1103/PhysRevLett.38.1440",
    journal = "Phys. Rev. Lett.",
    volume = "38",
    pages = "1440--1443",
    year = "1977"
}

@article{Peccei:1977ur,
    author = "Peccei, R. D. and Quinn, Helen R.",
    title = "{Constraints Imposed by CP Conservation in the Presence of Instantons}",
    reportNumber = "ITP-572-STANFORD",
    doi = "10.1103/PhysRevD.16.1791",
    journal = "Phys. Rev. D",
    volume = "16",
    pages = "1791--1797",
    year = "1977"
}

@article{Callan:1970ze,
    author = "Callan, Jr., Curtis G. and Coleman, Sidney R. and Jackiw, R.",
    title = "{A New improved energy - momentum tensor}",
    doi = "10.1016/0003-4916(70)90394-5",
    journal = "Annals Phys.",
    volume = "59",
    pages = "42--73",
    year = "1970"
}

@article{Ambjorn:1981xw,
    author = "Ambjorn, Jan and Wolfram, Stephen",
    title = "{Properties of the Vacuum. 1. Mechanical and Thermodynamic}",
    reportNumber = "CALT-68-855",
    doi = "10.1016/0003-4916(83)90065-9",
    journal = "Annals Phys.",
    volume = "147",
    pages = "1",
    year = "1983"
}

@article{Boehm:2003hm,
    author = "Boehm, C. and Fayet, Pierre",
    title = "{Scalar dark matter candidates}",
    eprint = "hep-ph/0305261",
    archivePrefix = "arXiv",
    doi = "10.1016/j.nuclphysb.2004.01.015",
    journal = "Nucl. Phys. B",
    volume = "683",
    pages = "219--263",
    year = "2004"
}

@article{Burgess:2000yq,
    author = "Burgess, C. P. and Pospelov, Maxim and ter Veldhuis, Tonnis",
    title = "{The Minimal model of nonbaryonic dark matter: A Singlet scalar}",
    eprint = "hep-ph/0011335",
    archivePrefix = "arXiv",
    reportNumber = "TPI-MINN-00-46, UMN-TH-1922-00, MCGILL-00-31, IASSNS-HEP-00-83",
    doi = "10.1016/S0550-3213(01)00513-2",
    journal = "Nucl. Phys. B",
    volume = "619",
    pages = "709--728",
    year = "2001"
}

@article{Barbieri:2006dq,
    author = "Barbieri, Riccardo and Hall, Lawrence J. and Rychkov, Vyacheslav S.",
    title = "{Improved naturalness with a heavy Higgs: An Alternative road to LHC physics}",
    eprint = "hep-ph/0603188",
    archivePrefix = "arXiv",
    reportNumber = "UCB-PTH-06-04, LBNL-59894",
    doi = "10.1103/PhysRevD.74.015007",
    journal = "Phys. Rev. D",
    volume = "74",
    pages = "015007",
    year = "2006"
}

@article{Dowker:1975tf,
    author = "Dowker, J. S. and Critchley, Raymond",
    title = "{Effective Lagrangian and Energy Momentum Tensor in de Sitter Space}",
    reportNumber = "Print-75-1015 (MANCHESTER)",
    doi = "10.1103/PhysRevD.13.3224",
    journal = "Phys. Rev. D",
    volume = "13",
    pages = "3224",
    year = "1976"
}

@article{Hawking:1976ja,
    author = "Hawking, S. W.",
    title = "{Zeta Function Regularization of Path Integrals in Curved Space-Time}",
    reportNumber = "PRINT-77-0293 (CAMBRIDGE)",
    doi = "10.1007/BF01626516",
    journal = "Commun. Math. Phys.",
    volume = "55",
    pages = "133",
    year = "1977"
}

@article{DeWitt:1975ys,
    author = "DeWitt, Bryce S.",
    title = "{Quantum Field Theory in Curved Space-Time}",
    doi = "10.1016/0370-1573(75)90051-4",
    journal = "Phys. Rept.",
    volume = "19",
    pages = "295--357",
    year = "1975"
}

@article{Blau:1988kv,
    author = "Blau, Steve and Visser, Matt and Wipf, Andreas",
    title = "{Zeta Functions and the Casimir Energy}",
    eprint = "0906.2817",
    archivePrefix = "arXiv",
    primaryClass = "hep-th",
    reportNumber = "MPI-PAE-PTH-18-88, LA-UR-88-1542",
    doi = "10.1016/0550-3213(88)90059-4",
    journal = "Nucl. Phys. B",
    volume = "310",
    pages = "163",
    year = "1988"
}

@article{Elizalde:1989dd,
    author = "Elizalde, E. and Romeo, A.",
    title = "{Heat Kernel Approach to the Zeta Function Regularization of the Casimir Energy for Domains With Curved Boundaries}",
    reportNumber = "UB-ECM-PF-3/89",
    doi = "10.1142/S0217751X90000751",
    journal = "Int. J. Mod. Phys. A",
    volume = "5",
    pages = "1653",
    year = "1990"
}

@book{Elizalde:1994gf,
    author = "Elizalde, E. and Odintsov, S. D. and Romeo, A. and Bytsenko, A. A. and Zerbini, S.",
    title = "{Zeta regularization techniques with applications}",
    doi = "10.1142/2065",
    isbn = "978-981-02-1441-8, 978-981-4502-98-6",
    publisher = "World Scientific Publishing",
    address = "Singapore",
    year = "1994"
}

@book{Elizalde:1995hck,
    author = "Elizalde, E.",
    title = "{Ten physical applications of spectral zeta functions}",
    doi = "10.1007/978-3-540-44757-3",
    isbn = "978-3-642-29404-4, 978-3-642-29405-1",
    volume = "35",
    year = "1995"
}

@article{Bordag:1996ma,
    author = "Bordag, Michael and Elizalde, E. and Kirsten, K. and Leseduarte, S.",
    title = "{Casimir energies for massive fields in the bag}",
    eprint = "hep-th/9608071",
    archivePrefix = "arXiv",
    reportNumber = "UB-ECM-PF-96-14",
    doi = "10.1103/PhysRevD.56.4896",
    journal = "Phys. Rev. D",
    volume = "56",
    pages = "4896--4904",
    year = "1997"
}

@article{Weinberg:1990rz,
    author = "Weinberg, Steven",
    title = "{Nuclear forces from chiral Lagrangians}",
    reportNumber = "UTTG-31-90",
    doi = "10.1016/0370-2693(90)90938-3",
    journal = "Phys. Lett. B",
    volume = "251",
    pages = "288--292",
    year = "1990"
}

@article{Machleidt:2011zz,
    author = "Machleidt, R. and Entem, D. R.",
    title = "{Chiral effective field theory and nuclear forces}",
    eprint = "1105.2919",
    archivePrefix = "arXiv",
    primaryClass = "nucl-th",
    doi = "10.1016/j.physrep.2011.02.001",
    journal = "Phys. Rept.",
    volume = "503",
    pages = "1--75",
    year = "2011"
}

@article{Mohr:2024kco,
    author = "Mohr, Peter J. and Newell, David B. and Taylor, Barry N. and Tiesinga, Eite",
    title = "{CODATA recommended values of the fundamental physical constants: 2022*}",
    eprint = "2409.03787",
    archivePrefix = "arXiv",
    primaryClass = "hep-ph",
    doi = "10.1103/RevModPhys.97.025002",
    journal = "Rev. Mod. Phys.",
    volume = "97",
    number = "2",
    pages = "025002",
    year = "2025"
}

@article{Adelberger:2003zx,
    author = "Adelberger, E. G. and Heckel, Blayne R. and Nelson, A. E.",
    title = "{Tests of the gravitational inverse square law}",
    eprint = "hep-ph/0307284",
    archivePrefix = "arXiv",
    doi = "10.1146/annurev.nucl.53.041002.110503",
    journal = "Ann. Rev. Nucl. Part. Sci.",
    volume = "53",
    pages = "77--121",
    year = "2003"
}

@book{Olver:2010ouy,
    editor = "Olver, Frank W. J. and Lozier, Daniel W. and Boisvert, Ronald F. and Clark, Charles W.",
    title = "{NIST Handbook of Mathematical Functions}",
    publisher = "Cambridge University Press",
    address = "New York, NY",
    isbn = "9780521140638",
    year = "2010"
}
\end{document}